\documentclass[vecphys,envcountchap]{svmult}

\usepackage{ergebnisse} % enlarge print area

\usepackage{makeidx}         % allows index generation
\usepackage{graphicx}        % standard LaTeX graphics tool
\usepackage{multicol}        % used for the two-column index
\usepackage[bottom]{footmisc}% places footnotes at page bottom
\makeindex             % used for the subject index
\begin{document}

%\chapter{1}

\title*{Quasars and their host galaxies}
\author{Mark Lacy}
%\and
%Name of Author\inst{2}}
\institute{}
%\institute{Spitzer Science Center, California Institute of Technology, MS220-6,
%Pasadena, CA 91125
%\texttt{mlacy@ipac.caltech.edu}}
%\and Name and Address of your Institute \texttt{name@email.address}}
\maketitle

%Your text goes here. Separate text sections with the standard \LaTeX\
%sectioning commands.

\begin{abstract}

This review attempts to describe developments in the fields of quasar and 
quasar host galaxies in the past five. In this 
time period, the Sloan and
2dF quasar surveys have added several tens of thousands of quasars, with 
Sloan quasars being found to $z>6$. Obscured, or partially obscured 
quasars have begun to be found in significant numbers.  
Black hole mass estimates for quasars, and our confidence in them, have 
improved significantly, allowing a start on relating quasar properties
such as radio jet power to fundamental parameters of the quasar such as
black hole mass and accretion rate. Quasar host galaxy studies have
allowed us to find and characterize the host galaxies of quasars to $z>2$.
Despite these developments, many questions remain unresolved, in particular 
the origin of the close relationship between black hole mass and galaxy
bulge mass/velocity dispersion seen in local galaxies.

\end{abstract}

\section{Introduction}
\label{sec:1}
% Always give a unique label
% and use \ref{<label>} for cross-references
% and \cite{<label>} for bibliographic references
% use \sectionmark{}
% to alter or adjust the section heading in the running head
%Your text goes here. Use the \LaTeX\ automatism for your citations
%\cite{monograph}.

Quasar astronomy has made some significant advances over the past 
few years. In this review we will focus on recent developments in the 
field, and in particular on the links between quasars, their black holes
and their host galaxies. 

The recent discovery that the mostly dormant black holes in the 
nuclei of nearby galaxies have masses which correlate with the luminosity
and velocity dispersion of their host galaxies has directly linked the
quasar phenomenon to galaxy evolution. Black hole mass estimates from
host galaxy luminosities can be compared to those which 
use the results of reverberation mapping
and the widths of broad emission lines, with generally consistent
results (Laor 1998; Onken et al.\ 2004). 
This gives us the possibility of understanding the 
black hole mass -- galaxy mass correlation through a study of the 
evolution of quasars and their host galaxies.

Reliable black hole mass estimates have also
stimulated long-standing debates on how the observational properties
of quasars, such as their emission-line spectra, radio-loudness and 
accretion rates, may (or may not) depend on black hole mass
(McLure et al.\ 1999; Laor 2000; Lacy et al.\ 2001; Boroson 2002; 
Woo \& Urry 2002). The correlation between black hole mass and 
the mass of the host galaxy have reinvigorated studies on the links 
between quasar activity and galaxy formation. In particular, 
the relationship between galaxy mergers, starbursts and AGN has
come under increased scrutiny using both results from 
Sloan Digital Sky Survey (SDSS) (Kauffmann et al.\ 2003) and
{\em Hubble Space Telescope (HST)} studies of quasar host galaxies 
(McLure et al.\ 1999; Canalizo \& Stockton 2000). 

A further recent development, 
the advent of large, uniform quasar surveys in the 
optical (the  SDSS quasar survey and Two Degree Field (2dF)
quasar survey) and the 
ROSAT X-ray survey, has also helped progress in the subject.
One problem that has bedevilled quasar astronomy throughout its history
is that selection of quasar samples has been very prone to 
selection effects. Optical
and (soft) X-ray selection techniques are only 
sensitive to quasars with little
dust or gas in the host galaxy to redden or absorb the quasar light. 
Radio selection is not sensitive to reddening, but only $\sim 10$\%
of quasars are bright radio emitters, and the selection effects associated
with radio quasar samples are only just beginning to be understood. 
In particular, the SDSS is 
able to pick objects that would be missed
from traditional quasar surveys (Fan 1999; Richards
et al.\ 2001), and also objects showing only narrow lines in the 
optical (Zakamska et al.\ 2003). Further red quasar selection using 2MASS, 2MASS/FIRST, 
hard X-ray and now Spitzer surveys is finally revealing the population of 
dust-shrouded type-2 quasars, whose existence has long been suspected, but
of which, until recently, there were few known examples.

Detection of the faint host galaxies of bright quasars has been
a long-standing problem in observational extragalactic astronomy. Although the
host galaxy and the bright, unresolved quasar nucleus typically have similar
total fluxes, the diffuse nature of the galaxy emission can frequently be 
confused with the extended wings of a poorly-characterized point spread 
function (PSF). 
Cosmological surface brightness dimming makes detecting the host galaxies
of high redshift quasars particularly challenging.  
Breakthroughs in quasar host
studies came with the advent of the 
{\em HST}, with its small, stable PSF, and
near-infrared array detectors. Early 
HST studies with the optical instruments on board allowed the detailed 
study of nearby quasar hosts (Disney et al.\ 1995; Bahcall et al.\ 1997). 
In parallel, ground-based studies in the near-infrared were able to study the
quasars at wavelengths where the flux of the quasar was minimized with 
respect to the flux of the host galaxy (Dunlop et al.\ 1993). The marriage
of $HST$'s small PSF and the near-infrared NICMOS detector 
allowed routine discoveries of quasar hosts 
up to $z\sim 2$ (Ridgway et al.\ 2001; Kukula et al.\ 2001). 

Adaptive optics are a relatively recent 
addition to the available techniques for quasar host imaging. Although
problematic in some respects (principally PSF variability) AO offers the
ability to study larger samples than are practical with the 
limited observing time available with HST, and, through using 
10m-class telescopes, 
better resolution and surface brightness sensitivity.

We end this review with a discussion of some of the remaining open
questions in quasar astronomy, and how advances in telescopes, instrumentation
and multiwavelength archives may be able to answer at least some of them in 
the years to come.

%In this review I will concentrate on developments in the past few years, 
%during which rapid progress has been made in several areas. These new 
%discoveries have, however, often raised more questions than they have 
%provided answers.

When calculating the intrinsic properties of the quasars 
we assume a cosmology with $H_0=70 {\rm kms^{-1}Mpc^{-1}}$, 
$\Omega_{\Lambda}=0.7$ and $\Omega_{\rm M}=0.3$. 
 
\section{New surveys for quasars}

Two major new quasar surveys with many tens of thousands of objects have recently appeared.
The 2dF quasar survey (Croom et al.\ 2004a)
uses photographic plate material to select UV-excess objects. It is thus a fairly
conventional survey, but, by using fibres from a large-scale survey of part of 
the southern sky conducted by the 2dF multifibre instrument on the 
Anglo-Australian Telescope  it has been able to 
find many thousands of objects. The SDSS quasar survey uses a multicolour technique to find
point-like objects off the stellar locus in colour space (Fan 1999, Richards et al.\ 2001). It is thus 
also able to find samples of lightly dust-reddened quasars and very high redshift ($z>2$)
objects which the UV excess technique cannot (Hall et al.\ 2002; Fan et al.\ 2004). 
The SDSS survey is not as deep as the 2dF, but covers a larger sky area with greater 
photometric accuracy. 
The Faint Images of the Radio Sky at Twenty-cm (FIRST) radio survey, a 
large area radio survey to mJy 
sensitivity, was combined with digitally-scanned photographic plate data to 
find quasar candidates, selected with a blue colour cut (but one much less 
severe than optically-selected surveys), to form the 
FIRST Bright Quasar Survey (FBQS, White et al.\ 1999). 
Although the FBQS is much smaller than
the SDSS and 2dF surveys, was interesting as it was the first survey
able to find significant numbers of quasars with 
radio fluxes between those of radio-loud and radio-quiet quasars.

The quasar luminosity function has been measured from both the 2dF and the SDSS surveys. 
Croom et al.\ (2004a) have analysed the 2dF quasar survey and find that a double power-law with
luminosity evolution provides a good fit to the data. Joint use of the 2dF and SDSS databases 
and techniques is now being applied to further refine estimates of
the luminosity function (Richards et al.\ 2005).

\section{Black hole masses and their implications}

Perhaps the biggest single advance in the study of quasars in recent years has been our 
increased confidence in estimates of masses of their black holes. 
Although a long-term campaign to obtain data for reverberation mapping resulted in 
estimates of the sizes of broad line regions (BLRs) (and hence, via linewidths and
assumptions about cloud dynamics, the masses of the central black holes), it was not until
black hole masses could be independently estimated by stellar and gas-dynamical techniques 
that it became common to use these estimates. Wandel (2002), Peterson et al.\ (2004) and 
Onken et al.\ (2004) showed that black 
hole masses derived from reverberation mapping and from bulge luminosities of Seyfert-1
galaxies were consistent, and Nelson et al.\ (2004) showed that black hole masses estimated from 
the velocity dispersions of Seyfert-1 galaxies were consistent with the reverberation
mapping results. Uncertainties of a factor of a few, e.g.\ the 
unknown broad-line region geometry, remain (Krolik 2001). Unfortunately, reverberation mapping
studies are restricted to a few tens of (mostly) low luminosity AGN. Thus the relationship 
between AGN luminosity and BLR radius is not very well established, and its extension to 
the high luminosities of high redshift quasars uncertain. Reverberation mapping studies
use H$\beta$, whereas studies of high redshift quasars are usually restricted to UV
lines which are redshifted into the optical such as Mg{\sc ii} and C{\sc iv}
(McLure \& Jarvis 2002; Vestergaard 2002). Despite these problems and 
uncertainties, the ability to obtain even order-of-magnitude black hole 
mass estimates has allowed some interesting studies to be made.

A further technique for measuring black hole mass was suggested by Nelson (2000), using the 
width of narrow [O{\sc iii}] emission line as a proxy for the bulge velocity dispersion. 
This seems to work well for radio-quiet quasars, but not for radio-louds, 
where the narrow-line region dynamics are frequently affected by shocks induced by 
the radio jets. This technique is easy to apply at moderate redshifts, and near-infrared
spectra can be used at higher redshifts (Shields et al.\ 2003).  
The discovery of significant blueshifts in the [O{\sc iii}] line relative to systemic 
velocities (Boroson 2005) has cast doubt on the accuracy of this technique, however. 

By combining black hole masses and measured luminosities, we have another key physical 
parameter, namely accretion rate relative to the Eddington Limit. Studies seem
to indicate that the Eddington Rate is indeed a hard upper limit to the accretion 
rate (e.g.\ McLure \& Dunlop 2004), with a typical optically-selected quasar having an 
Eddington Ratio (i.e.\ the ratio of 
accretion rate relative to the Eddington Limit) $\sim 0.1$. One important 
component to this calculation is the bolometric correction factor to 
convert quasar luminosities in a given waveband to total accretion 
luminosities. Most studies still use the Elvis et al. (1994) corrections, but 
new observations of mostly fainter quasars, in particular with the Spitzer Space Telescope in 
the mid-infrared and the Galaxy Evolution Explorer (GALEX) in the ultraviolet 
should enable us to significantly improve these corrections and allow luminosity and
redshift effects to be fully taken into account.

Estimates for black hole masses have been used to relate observational correlations to 
physical properties of the AGN. Boroson (2002) showed that the 
eigenvector 1 of Boroson \& Green (1992) is related to the Eddington Ratio. 
Objects with high accretion rates tend to be at one
extreme, with high Fe{\sc ii}  emission and low [O{\sc iii}] emission, 
while those with 
low accretion rates are at the other end. 

\section{Radio-loudness of quasars}

Another area in which black hole mass estimates have been used to try to improve our 
physical understanding of AGN is that of quasar radio emission.
Amongst well-defined samples of quasars there is a weak, but 
significant relation between black hole mass and radio-loudness in the sense that the
radio-loud quasars have higher mean black holes masses [Laor (2000), Lacy et al.\ (2001), Boroson
(2002), Jarvis \& McLure (2002), Metcalf \& Magliochetti (2005)]. 
Studies of more heterogeneous 
samples fail to find a correlation of radio-loudness with black hole mass, 
however (Woo \& Urry 2002). When low-luminosity
AGN are included amongst the samples being studied, 
an anti-correlation of radio-loudness with accretion rate relative to the 
Eddington Limit is seen (Ho 2002), albeit with a lot of scatter. 
Whether or not the correlation of black hole mass and radio-loudness is  
real it is surprising that the relationship 
between radio emission and the physical characteristics of the quasar (accretion rate 
and black hole mass) is so hard to pin down. Observations of X-ray binary 
``microquasars'' show X-ray emission properties that change significantly during a radio 
outburst (a hardening of the X-ray spectrum, hypothesized to be due to the 
temporary formation of an ion-supported torus in the inner part of the accretion 
disk and a softening during its subsequent collapse, e.g.\ Fender, Belloni \& Gallo 2004), 
but no obviously analogous behavior is seen in 
radio-loud quasars. This may be partly a question of timescales of course, but even so 
one might expect a much bigger difference in the observed properties of radio-loud and
radio-quiet quasars were the microquasar analogy to be followed exactly. One promising 
analogy is between FRI radio galaxies and microquasars in the ``low/hard'' state
with advection-dominated accretion flows, which are able to produce strong outflows (and perhaps radio jets) 
very efficiently (e.g.\ Blandford \& Begelman 1999). 
However, directly applying this analogy
to high accretion rate radio-loud quasars, which appear to have both highly-efficient relativistic jet
production and classic accretion disks, fails.  

Black hole spin has long been suspected as being instrumental in radio-loudness. In 
particular, the 
Blandford-Znajek mechanism (Blandford \& Znajek 1977), 
in which jets are produced by magnetic fields threading the 
ergosphere of the black hole requires a rapidly spinning black hole whose
rotational kinetic energy is tapped to power the radio jets. 
X-ray spectroscopy of the relativistic
Fe K$\alpha$ line in Seyfert galaxies can, in principle, be used to measure the
radius of the last stable orbit, and hence the spin of the black hole.
This is just possible 
with XMM and Chandra, but to date only one object, MGC-6-30-15, has been studied in 
sufficient depth to fit the Fe K$\alpha$ line. Unfortunately, uncertainties in the underlying 
continuum emission and possible absorption features 
mean that a definitive statement about whether or not the black hole is 
spinning is hard to make. Nevertheless, a model with a spinning black 
hole is preferred by Vaughan \& Fabian (2004). Unfortunately for
the proponents of the spinning black hole model for the production of radio 
jets this Seyfert galaxy is radio-quiet.

Closely related to the question of the mechanism for radio-loudness is the question of whether
there exists a dichotomy in the radio-loudness parameter (i.e.\ the ratio of radio to optical/UV
luminosity). Early
studies of the PG quasar sample suggested this was the case, but more recent surveys
have been more ambiguous. For example, the FBQS shows no 
dichotomy (e.g.\ Lacy et al.\ 2001). 
The case for dichotomy in the much larger SDSS and 2dF quasar samples is probably not made. 
Although Ivecic et al. (2002) claim that one exists in the SDSS, the claim is based on indirect
evidence. A similar analysis by Cirasuolo et al.\ (2003) based largely on 2dF quasars
showed no evidence for a dichotomy. It is probably fair 
to say, however, that pending deep, large-area radio surveys, the
jury is still out on both whether there is a correlation of radio-loudness with 
black hole mass and whether there is a dichotomy in the 
distribution of radio-loudness.

\section{Clustering of quasars}

One interesting product of the large-scale quasar surveys is an estimate of quasar clustering
and its evolution with redshift. Clustering statistics give information on the masses
of the dark matter haloes containing the quasars and the value of the bias parameter (the 
extent to which objects fail to trace the underlying mass distribution). Croom et al.\ (2005)
use the 2dF survey to analyse quasar clustering and its evolution using a sample of over 
20000 quasars. They find that quasars occupy dark matter haloes with masses of a few $\times 
10^{12} M_{\odot}$ at all redshifts. 
An increase in the clustering with redshift is seen, 
ascribed to an increase in the bias parameter with redshift from close to unity at $z\approx 0$
to about a factor of four at $z\sim 2$.
Although quasars at $z\approx 0$ are seen in haloes 
with the space density of $L^*$ galaxies today, the $\sim 10^{12} M_{\odot}$ haloes
hosting quasars where rarer in the past, and the haloes hosting high-$z$ quasars 
will grow into group and cluster-sized haloes by the present epoch. Croom et al.\
interpret this as a trend for lower-mass black holes being more active at low redshift. 
Of course, in any flux limited survey, the characteristic luminosity of the objects will 
correlate strongly with redshift, and the possibility of an increase of clustering 
with luminosity, although not seen in the 2dF dataset, remains a viable alternative
explanation. Results from the SDSS, with its higher flux limit, should help to 
determine whether luminosity or evolution effects are responsible for these 
results.

\section{High-$z$ quasars and reionization}

The SDSS has allowed the efficient selection of several $z\sim 6$ quasars. These are 
important from a cosmological viewpoint for several reasons. The most distant
of these show Gunn-Peterson troughs, indicating that at least a small fraction  
of the intergalactic medium was neutral at $z\approx 6.5$ (Becker et al.\ 2002; 
Fan et al.\ 2003; White et al.\ 2003). Exactly how much is still 
the subject of debate, and larger samples of high-$z$ quasars will be needed 
to establish this (Mesinger, Haiman \& Cen 2004). 
The mere existence of $\sim 10^9 M_{\odot}$ black holes at $z\sim 6$ 
is itself interesting, particularly as the detection of Fe{\sc ii} emission lines
in near-infrared spectra shows that the 
ISM of the host was already significantly enriched with an element which, at least
locally, is mostly produced in type-1 supernovae, with a long ($\sim 0.3$Gyr) delay between 
star formation and the supernova explosions (Freudling et al.\ 2003; Barth et al.\ 2003). 
This implies an early epoch for the first major burst of  star formation in the
host galaxy ($z\stackrel{>}{_{\sim}} 10$). 
The number density of very high-$z$ quasars can be used to estimate their contribution
to reionization. Based on the numbers found so far, it seems that 
there are not enough of these objects to reionize the Universe (unless the faint end of the
quasar luminosity function is very steep), meaning that star-forming galaxies are the most 
likely culprits. 

\section{Broad absorption line quasars}

The past few years have seen significant developments in the study of 
broad-absorption line quasars (BALs), the quasar population which shows 
absorption by an ionized wind with velocities up to $0.1c$. The number 
of iron low-ionization BALs (FeLoBALs) known has increased significantly, 
due principally to new quasar surveys. The FBQS was the first to find 
significant numbers of these 
objects (Becker et al.\ 2000). Subsequently, several FeLoBALs
were found in the SDSS survey using colour-based selection criteria
(Hall et al.\ 2002). These remarkable objects 
have line blanketing by FeII shortward of $\sim 2800$\AA, frequently 
almost completely extinguishing the UV light from the quasar. They are 
also significantly reddened by dust, the combination
giving them quite different colours than normal quasars.

The discovery of a number of truly radio-loud BALs (including a couple with the
classic FRII morphology [Gregg et al.\ 2000; Brotherton et al.\ 2002]) has also 
disproved a long-standing suggestion that radio-loud quasars are incapable of hosting BAL flows.
Further radio studies of the FBQS BALs has also undermined the popular
orientation model for BALs, in which the BAL winds arise from the surface
of the accretion disk, with the line of sight passing through them. This model 
predicts that BAL quasars are all seen with their accretion disks nearly edge-on.
Becker et al.\ (2000) showed that
the radio spectral indices of BALs range from steep to flat, indicating
no special orientation for the BAL population.

Modelling of the gas flows causing broad absorption lines has improved
significantly, with careful allowance being made for saturation and partial 
covering effects (e.g.\ Arav et al.\ 2001). Attempts to model the very 
low ionization outflows in the FeLoBAL quasars have raised interesting 
questions as to whether the low ionization absorbing gas is very far out, 
at distances of hundreds of parsec (de Kool et al.\ 2002), 
or whether the outflow is comprised of a multi-phase gas with dense, 
low-ionization clouds
embedded in a low density, high ionization wind (Everett, K\"{o}nigl \& Arav
2002).

\section{Finding the obscured quasar population}

Recent years have also shown great advances in finding the obscured quasar population. 
Initially, progress was slow. In the hard X-ray, the small field sizes and depth required
to find the obscured quasar population restricted discoveries to a few objects 
(e.g.\ Norman et al.\ 2002, Stern et al.\ 2002, Padovani et al.\ 2004). 
The 2MASS near-infrared survey was able to find some dust-reddened,
mostly low-$z$ quasars (Cutri et al.\ 2002) and by combining 2MASS with the FIRST radio survey, 
several higher redshift, radio-intermediate 
reddened quasars were found (Glikman et al.\ 2004). However, 
2MASS selection was unable to find objects with rest-frame optical 
reddenings ($A_V$) greater than a few magnitudes. 

Finding the heavily obscured objects, which show no broad 
lines in the optical and which have rest-frame $A_V\sim 10-100$, has proven difficult, 
and it is only very recently that significant 
progress has been made.  
Zakamska et al.\ (2003) were able to isolate moderate
redshift objects with high-ionization narrow lines characteristic of type-2 AGN in the SDSS survey, 
and later 
showed that many of these objects show broad lines in polarized light, confiming their
quasar nature. A combination of mid-infrared fluxes from the {\em Infared Space Observatory
(ISO)} and 2MASS was able to find
objects significantly more reddened than those found by 2MASS alone 
(Leipski et al.\ 2004,2005).
Using mid-infrared colours from $Spitzer$, Lacy et al.\ (2004, 2005) also found 
examples of this type-2 quasar population. A strategy of searching for 
$Spitzer$-detected radio-intermediate objects has been used successfully by Martinez-Sansigre et 
al.\ (2005), who have found several $z>2$ type-2 quasars.
Selection effects for all these techniques still
need to be fully understood, but it is clear that the type-2 quasar population is at least
comparable in number density to the ``normal'' quasar population, and 
perhaps 2-3 times larger. 
Figure 1 shows the spectra of a type-1 quasar, a 
lightly-reddened FeLoBAL quasar and a heavily obscured type-2 quasar normalised to 
approximately the same emission line fluxes.

An upper bound on the total number of quasars, both obscured and unobscured, can be
obtained by comparing the mass in black holes in the centers of galaxies today with the total 
accretion luminosity 
of the quasar population, divided by the 
accretion efficiency, a point orignally 
made by Soltan (1982). The most up-to-date versions of this calculation (e.g. Yu \& 
Tremaine 2002) leave little room for a very large obscured quasar population, but 
uncertainties about the exact value of the accretion efficiency (i.e. the ratio of the 
mass-energy accreted onto the black hole to that liberated as accretion luminosity) still 
allow a significant obscured (or indeed advection-dominated accreting) population. In the 
standard accretion disk model, accretion efficiency is 
determined primarily by the radius of the last stable orbit. For a non-spinning 
(Schwartzschild) black hole this limits the accretion efficiency to $\sim 0.1$, but the 
frame-dragging effect of a spinning (Kerr) black hole allows for a closer-in last stable 
orbit and accretion efficiencies $\sim 0.3$. Efficiencies of this order are 
required to accommodate a significant hidden quasar population.

\begin{figure}
\centering
\includegraphics[height=10cm]{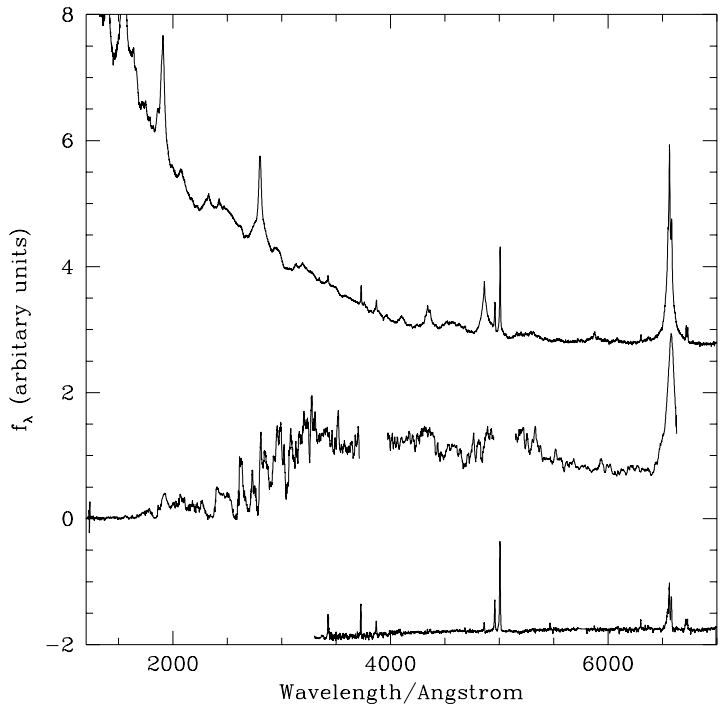}
%\picplace{5cm}{2cm} % Give the correct figure height and width in cm
%
\caption{Rest-frame spectra of traditional and non-traditional quasars: top - composite of
normal quasars from the SDSS (Vanden Berk et al.\ (2001); middle - the $z=2.65$ 
FeLoBAL quasar FTM J100424.9+122922 (Lacy et al.\ 2002a); bottom - a composite
of type-2 quasars from Spitzer selection from 3200 - 7000\AA$\;$
(Lacy et al.\ 2005). The top and
bottom spectra have been offset by $\pm 2$ units for clarity. All three 
spectra have been normalized to approximately the same emission line strength.
(The blank 
regions in the middle spectrum at $\approx 3800$ and $\approx 5100$\AA$\;$
are zero transparency regions in the near-infrared.)}
\label{fig:1}       % Give a unique label
\end{figure}

\section{Quasar host galaxy studies}

Although ground-based studies of quasar hosts have been proceeding for many years, it required
the advent of $HST$ to allow robust determinations of the host galaxy properties. 
Figure 1 shows the regions of the absolute magnitude vs redshift plot covered by the
major $HST$ programs.
WFPC2 observations of nearby quasars showed that luminous quasars are hosted
by massive galaxies, usually early types (Disney et al.\ 1995; Bahcall et al.\ 1997; 
McLure et al.\ 1999; Dunlop et al.\ 2003; Floyd et al.\ 2004). The advent of  
the NICMOS camera on HST allowed the study of rest-frame optical 
emission from host galaxies of quasars at high redshifts for the first time. Kukula et al.\
(2001) and Ridgway et al. (2001) showed that, at $z\sim 2$, the trend for 
quasars to be hosted in massive galaxies continues, though
less-luminous quasars are generally found in less-luminous hosts (Figure 2). 
The sample of Ridgway
et al.\ contained quasars of similar luminosity to the low-redshift quasars of 
Dunlop et al., allowing a direct comparison of the stellar masses of the hosts. This 
indicated that the stellar masses of the high-$z$ quasars were smaller by a factor 
of a few than their low-$z$ counterparts. Semi-analytic modelling by Kauffmann \& Haehnelt (2000) 
suggested that the hosts of high-$z$ quasars should be lower in stellar mass, though 
by a larger factor. Whether this difference is significant or not remains to be 
established, though the results of Kukula et al.\ on the hosts of
slightly more luminous quasars 
are in more disagreement with the predictions of that model. More recent 
semi-analytic models,
which include feedback effects from both supernovae and the AGN, and a different
prescription for star formation in massive dark matter haloes at early epochs agree
better with observations (e.g.\ Granato et al.\ 2003), though a robust, predictive model
for galaxy and quasar formation is still lacking. Figure 2 also shows that radio-loud 
quasars tend to be in more massive hosts (at a given quasar luminosity) than radio-quiets
at all redshifts (e.g. Kukula et al.\ 2001)

\begin{figure}
\centering
\includegraphics[height=10cm]{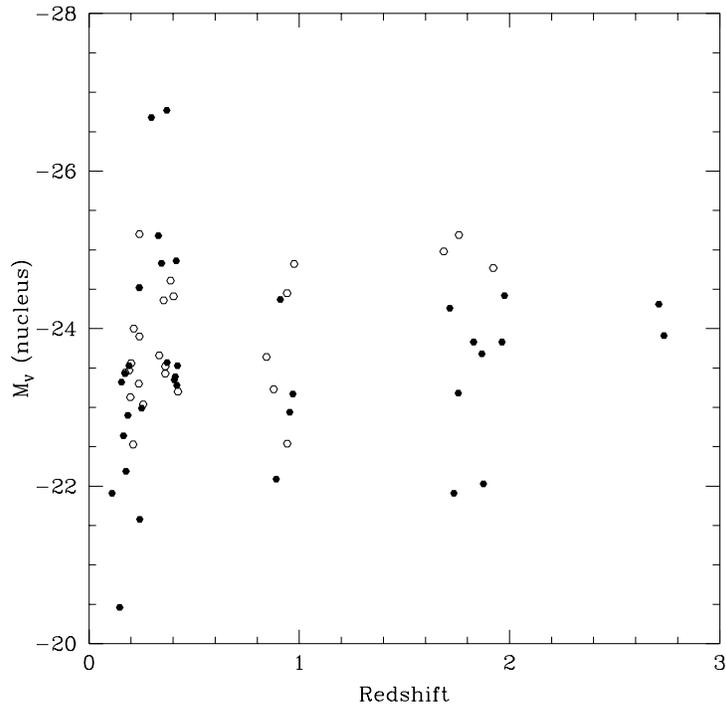}
%\picplace{5cm}{2cm} % Give the correct figure height and width in cm
%
\caption{The absolute-magnitude -- redshift plane for quasars studied by the largest 
$HST$ programs at $z>0.5$ (Kukula et al.\ 2001; Ridgway et al.\ 2001), and the two largest 
programs at low-$z$ (Dunlop et al.\ 2003; Floyd et al.\ 2004). All these
samples were imaged close to the rest-frame $V$-band. Although moderate luminosity 
quasars are well-represented at most redshifts, there is a lack of high-luminosity 
quasars, partly for practical reasons of the difficulty of PSF subtraction from quasars
whose nuclear light is significantly brighter than that from its host galaxy.
Radio-loud quasars are shown as open symbols, radio-quiets as filled symbols.}
\label{fig:2}       % Give a unique label
\end{figure}

\begin{figure}
\centering
\includegraphics[height=10cm]{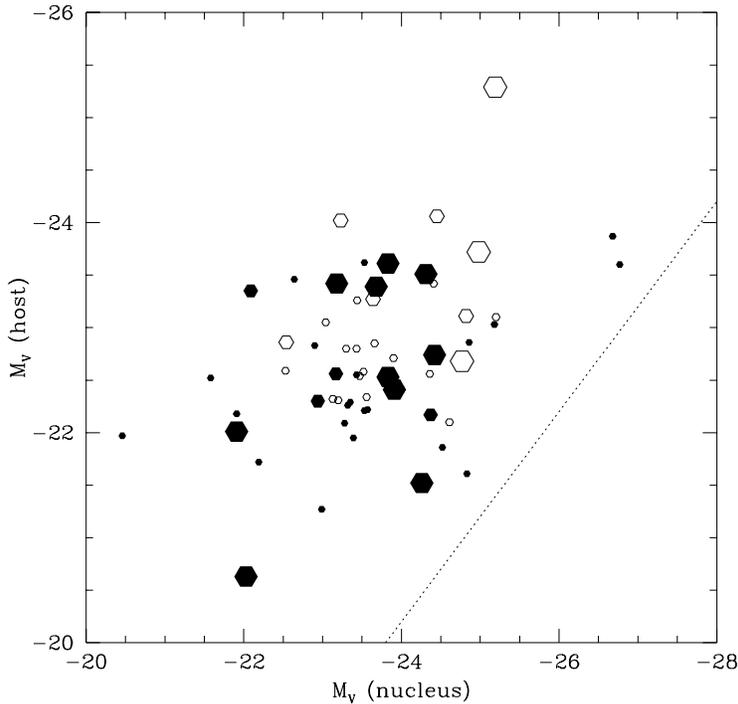}
%\picplace{5cm}{2cm} % Give the correct figure height and width in cm
%
\caption{The samples of Figure 2 with absolute magnitude of the host plotted against
nuclear absolute magnitude. Small symbols indicate quasars with $z<0.5$, medium, those with 
$0.5<z<1.5$, and large, those with $z>1.5$. The dotted line corresponds 
approximately to the luminosity expected from Eddington-limited
accretion [using the van der Marel (1999) relation]. In general, host and 
quasar luminosity are correlated, even though few quasars closely approach the Eddington Limit.
Note also that there is no strong systematic trend for
high-$z$ hosts to be fainter than their low-$z$ counterparts [as predicted, 
for example, by Kauffmann \& Haenelt (2000)], but the dispersion 
in host galaxy magnitudes does seem to be higher at $z\sim 2$, indicating that a
wider range of black hole masses may be involved in quasar activity at early epochs. 
(It should also
be borne in mind that the mass:light ratios of young stellar populations are likely
to be lower at high-$z$, so their black holes may be less massive than their position
on this plot would indicate.) }
\label{fig:3}       % Give a unique label
\end{figure}

Spectroscopy of quasar host galaxies has been attempted from the ground. Nolan et al.\ (2001)
found that most quasar host galaxies had evolved stellar populations, $\sim 10$Gyr
old, with only a very small amount of recent star formation. Deeper spectroscopy 
with Keck by Canalizo \& Stockton (2001), however, found evidence of bursts of star 
formation within the past $\sim 100$Myr in quasars 
with far-infrared excesses. Indeed, such populations have also been found in 
radio galaxies (Tadhunter et al.\ 2005), and in composite 
spectra of Seyfert galaxies and
low-luminosity quasars from the SDSS (Kauffmann et al.\ 2003). 
There is much debate about whether the 
different results are due to better quality spectra taken closer to the nucleus 
using 8-10m-class telescopes, or whether there are real differences in the 
host galaxy properties of the different quasar samples studied.

In addition to space-based observations, ground-based observations with adaptive optics 
(AO) are beginning to be exploited for quasar host galaxy studies (e.g., 
Stockton, Canalizo \& Close 1998, Lacy et al.\ 2002b, 
Croom et al.\ 2004b; Falomo et al.\ 2005). The high spatial resolution of 
AO is able to concentrate the light of the nucleus, leaving the host much more visible.
Although the PSF is not as stable as that produced by space-based observations, the relative
cheapness of ground-based telescope time (particularly on smaller telescopes) allows for 
large surveys not practical with space-based observatories, and on 8m-class telescopes, 
the resolution exceeds that from HST by a significant factor. Although it
is important to bear in mind that surface brightness sensitivity is 
fundamental to the detection of quasar hosts, AO does 
significantly improve the ability to detect hosts through concentrating the 
quasar light into the center of the galaxy image.With laser guide stars
becoming available at several major observatories, coupled with large quasar surveys 
which allow selection of quasars near bright guide stars, this area of research should 
have a significant impact in the years to come. Host galaxy studies with 8m-class telescopes
without adaptive optics, or with only active optics have also been attempted. Although
lacking in spatial resolution, the good surface brightness sensitivity of these
telescopes has allowed detections of host galaxies out to $z>4$ (e.g.\ Falomo et al.\ 2001; 
Hutchings 2003). 

An early motivation for quasar host galaxy studies was to search for the trigger mechanism
of quasar activity (e.g.\ Stockton \& MacKenty 1987). 
Mergers, or other galaxy interactions have long been thought of 
as likely triggers as they are able to disrupt stable gas which might fall onto the black 
hole. What has never been clear, however, is how disturbances due to mergers or interactions
on kpc scales influence the gas motions within the sub-pc inner region of the AGN. Indeed, 
McLure et al.\ (1999) conclude that few of their quasar hosts show obvious signs of interaction. 
However, the presence of moderate age stellar populations in quasar hosts suggests there
may be a delay between the large-scale disruption (which would trigger a starburst almost 
instantaneously) and the onset of quasar activity. One possibility is that dust-reddened
and/or type-2 quasars may represent an early stage in quasar evolution. 
Canalizo \& Stockton (2001) find that several of their IR-excess quasars are 
low-ionization BALs with significant 
dust reddening, and, although many type-2
quasars have classic high-ionization emission lines in their optical spectra, 
there are several whose spectra either lack emission lines, 
or contain only emission lines from starbursts (e.g.\ the
``type-3'' quasars of Leipski et al.\ 2005).
These might represent quasars in a very early, completely dust-shrouded 
phase in their evolution where even the narrow-line region is
obscured. On the other hand, an HST-based study of the host 
galaxies of 2MASS-selected dust-reddened QSOs by Marble et al.\ (2003)
found no significant difference between the hosts of dust-reddened and normal QSOs
in terms of the spread in galaxy type and evidence of recent merger activity. 

\section{Open questions}

Over the past few years quasar astronomy has undergone something of a revolution. From 
surveys of a few hundred quasars we have gone to surveys of several tens of thousands. We 
have much more confidence in our ability to estimate the important physical parameters
of quasars, namely black hole mass and accretion rate. The highest redshift quasar is now
at $z>6$. We have resolved the host galaxies of quasars out to $z>2$. 
We are close to being able to place much-improved constaints on the numbers of dust-hidden
quasars, which studies suggest at least equal, and perhaps exceed, the numbers of 
``classical'' blue quasars. Although finding the most heavily obscured objects 
will always be hard, nature seems to be helping us by providing many of these 
objects with the ``obscuring torus'' geometry which makes them much more visible 
in the mid-infrared. Nevertheless, we still
have open a number of important questions - indeed, our new information has allowed us to 
ask new ones. Perhaps the most fundamental is the nature and evolution of the black hole
mass -- bulge mass relation, as it demonstrates a clear link between quasars and their 
host galaxies. This link might be established by a mechanism in which the quasar
acts as a govenor providing negative feedback to star-forming processes in the host 
galaxy (e.g.\ Silk \& Rees 1998).

Other questions remain. Why does the evolution of quasars so closely resemble pure luminosity 
evolution, even though we know high redshift and low redshift quasars must be very 
different phenomena, one fundamentally linked to galaxy formation and growth, and the other
apparently a rare phenomenon which seems to at best reflect the end results of a merger several 
hundreds of Myr earlier? How are outflow phenomena such as radio-loud quasars and BALs
linked to the normal quasar population? Is the obscured quasar population linked to the
normal quasar population through evolution, orientation, or both? Exactly how many quasars 
are hidden by dust?

Future techniques to study quasars will probably involve a shift from the study of pre-selected quasar 
samples towards survey-based samples. Data mining of public multiwavelength surveys will 
result in the discovery of objects which occupy unusual parts of parameter space, such as 
high redshift type-2 quasars (Padovani et al.\ 2004). New radio telescopes such as the 
Low Frequency Array (LOFAR) and the Square Kilometer Array, which can image deep enough over a 
large enough area  to measure the radio-luminosity 
function of radio-quiet quasars will shed new light on the relationship
of radio jet power to other quasar properties. With robotic telescopes becoming more 
common, it becomes feasible to undertake reverberation mapping studies of larger samples
of objects for longer periods, including higher luminosity AGN and quasars, which can be used
to improve black hole mass estimates. Statistical studies of quasar
variability (e.g.\ Vanden Berk 2004; de Vries et al.\ 2005) will become easier and 
more accurate with e.g.\ the Large Synoptic Survey Telescope, and should 
open up new ways of investigating the quasar phenomenon. 

With the future of $HST$ uncertain, 
adaptive optics is certain to become a more important tool for the study of quasar host 
galaxies. Future space missions such as the James Webb Space Telescope and the Terrestrial
Planet Finder (Cononograph) will play an important role, but both missions 
are some years away. Of the open questions, probably only the nature and numbers of the 
obscured quasar population is one we will be able to answer soon. The others will require
further observations and theoretical modelling before we will be able to make progress.

\section*{Acknowledgements}

I would like to thank Susan Ridgway for reading through the manuscript, and
my other long-term collaborators on studies of quasars: Bob Becker,
Gabriela Canalizo, Elinor Gates, Michael Gregg, Steve Rawlings and Margrethe Wold, 
discussions with whom over the years have helped to shape this review.

\bibliographystyle{}
\bibliography{}

\begin{thebibliography}{}
\bibitem[]{} Arav, N.\ et al.\ (1999), Hubble Space Telescope Observations of the Broad Absorption Line Quasar PG 0946+301, ApJ, 516, 27
\bibitem[]{} Bahcall, J.N., Kirhakos S., Saxe, D.H.\ \& Schnieder,D.P.\ 
(1997) Hubble Space Telescope Images of a Sample of 20 Nearby Luminous 
Quasars, ApJ, 479, 642
\bibitem[]{} Barth, A.J., Martini, P., Nelson, C.H.\ \& Ho, L.C.\ (2003) Iron Emission in the $z = 6.4$ Quasar SDSS J114816.64+525150.3, ApJ, 594, L95
\bibitem[]{} Becker, R.H., White, R.L., Gregg, M.D., Brotherton, M.S., Laurent-Muehleisen, S.A., Arav, N.\ (2000) Properties of Radio-selected Broad Absorption Line Quasars from the First Bright Quasar Survey, ApJ, 538, 72
\bibitem[]{} Becker, R.H.\ et al.\ (2001) Evidence for Reionization at 
$z \sim 6$: Detection of a Gunn-Peterson Trough in a $z=6.28$ Quasar, AJ, 122, 2850
\bibitem[]{} Blandford, R.D.\ \& Begelman, M.C.\ (1999) On the fate of gas accreting at a low rate on to a black hole, MNRAS, 303, L1
\bibitem[]{} Blandford, R.D.\ \& Znajek, R.L.\ (1977) Electromagnetic extraction of energy from Kerr black holes, MNRAS, 179, 433
\bibitem[]{} Boroson, T.A.\ \& Green, R.F. (1992) The emission-line properties of low-redshift quasi-stellar objects, ApJS, 80, 109
\bibitem[]{} Boroson, T.A., (2002) Black Hole Mass and Eddington Ratio as Drivers for the Observable Properties of Radio-loud and Radio-quiet QSOs, ApJ, 565, 78
\bibitem[]{} Boroson, T.\ (2005) Blueshifted [O III] Emission: Indications of a Dynamic NLR, AJ, 130, 381
\bibitem[]{} Brotherton, M.S., Croom, S.M., de Breuck, C., Becker, R.H.\ \& Gregg, M.D.\ (2002) The 
Twice-Overlooked, Second Fanaroff-Riley II Broad Absorption Line Quasar LBQS 1138-0126, AJ< 124, 2575
\bibitem[]{} Canalizo, G.\ \& Stockton, A.N.\ (2001) Quasi-Stellar Objects, Ultraluminous Infrared Galaxies, and Mergers, ApJ, 555, 719
\bibitem[]{} Cirasuolo, M.\ Celotti, A., Magliocchetti, M., Danese, L.\ (2003) Is there a dichotomy in the radio loudness distribution of quasars?, MNRAS, 346, 447
\bibitem[]{} Croom, S.M.\ et al.\ (2004a) The 2dF QSO Redshift Survey -- XIV. Structure and 
evolution from the two-point correlation function, MNRAS, 356, 415
\bibitem[]{} Croom, S.M., Schade, D., Boyle, B.J., Shanks, T., Miller, L.\ \&
Smith, R.J.\ (2004b) Gemini Imaging of QSO Host Galaxies at $z\sim 2$, ApJ, 606, 126
\bibitem[]{} Cutri, R.M., Nelson, B.O., Francis, P.J.\ \& Smith, P.S.\ (2002), The 2MASS Red AGN Survey, 
in, AGN Surveys, Proceedings of IAU Colloquium 184, ASP Conference Proceedings, Vol. 284. Edited by R.F.
Green, E.Ye. Khachikian, and D.B. Sanders. San Francisco, CA: Astronomical Society of the Pacific, p. 127.
\bibitem[]{} de Kool, M.\ et al., (2002), Keck HIRES Spectroscopy of the 
Fe II Low-Ionization Broad Absorption Line Quasar FBQS 0840+3633: Evidence 
for Two Outflows on Different Scales, ApJ, 570, 514
\bibitem[]{} de Vries, W., Becker, R.H., White, R.L.\ \& Loomis, C.\ (2005), Structure Function Analysis of Long-Term Quasar Variability, AJ, 129, 615
\bibitem[]{} Disney, M.J.\ et al.\ (1995) Interacting Elliptical Galaxies 
as Hosts of Intermediate Redshift Quasars, Nat, 376, 150
\bibitem[]{} Dunlop, J.S., Taylor, G.L., Hughes, D.H.\ \& Robson, E.I.\ 
(1993) A near-IR study of the host galaxies of radio-quiet quasars, 
radio-loud quasars and radio galaxies, MNRAS, 264, 455 
\bibitem[]{} Dunlop, J.S.\ et al.\ (2003) Quasars, their host galaxies and their central black holes, MNRAS, 340, 1095
\bibitem[]{} Elvis, M.\ et al.\ (1994) Atlas of quasar energy distributions, ApJS, 95, 1
\bibitem[]{} Everett, J.\ K\"{o}ngl, A.\ \& Arav, N.\ (2002), 
Observational Evidence for a Multiphase Outflow in Quasar FIRST J1044+3656,
ApJ, 569, 671
\bibitem[]{} Falomo, R.\ et al.\ (2001) Near-Infrared Imaging of the Host Galaxies of Three Radio-loud Quasars at $z\sim 1.5$, ApJ, 547, 124
\bibitem[]{} Falomo, R.\ et al.\ (2005) VLT adaptive optics imaging of QSO host galaxies and their close environment at $z \sim 2.5$: Results from a pilot program, A\&A, 434, 469
\bibitem[]{} Fan, X.\ (1999) Simulation of Stellar Objects in SDSS Color 
Space, AJ, 117, 2528 
\bibitem[]{} Fan, X.\ et al.\ (2003) Evolution of the Ionizing Background and the Epoch of Reionization from the Spectra of z~6 Quasars, AJ, 123, 1247
\bibitem[]{} Fan, X.\ et al.\ (2004) A Survey of $z>5.7$ Quasars in the Sloan Digital Sky Survey. III. Discovery of Five Additional Quasars, AJ, 128, 515
\bibitem[]{} Fender, R.P., Belloni, T.N.\ \& Gallo, E.\ (2004) Towards a unified model for black hole X-ray binary jets, MNRAS, 355, 1105
\bibitem[]{} Floyd, D.J.E., et al.\ (2004) The host galaxies of luminous quasars, MNRAS, 355, 196
\bibitem[]{} Freudling, W., Corbin, M.R.\ \& Korista, K.T.\ (2003) Iron Emission in z~6 QSOS, ApJ, 587, L67
\bibitem[]{} Granato, G.L.\ et al.\ (2003) A Physical Model for the Coevolution of QSOs and Their Spheroidal Hosts,
ApJ, 600, 580
\bibitem[]{} Gregg, M.D., Becker, R.H.\ Brotherton, M.S., Laurent-Muehleisen, S., Lacy, M.\ \& White, R.L.\ (2000) Discovery of a Classic FR II Broad Absorption Line Quasar from the FIRST Survey, ApJ, 544, 142
\bibitem[]{} Hall, P.B.\ et al.\ (2002) Unusual Broad Absorption Line Quasars from the Sloan Digital Sky Survey, ApJS, 141, 267
\bibitem[]{} Ho, L.\ (2002) On the Relationship between Radio Emission and Black Hole Mass in Galactic Nuclei, ApJ, 564, 120
\bibitem[]{} Hutchings, J.B., (2003), Host Galaxies of $z \sim 4.7$ Quasars, AJ, 125, 1053
\bibitem[]{} Glikman, E.\ et al.\ (2004) FIRST-2Mass Sources below the APM Detection Threshold: A Population of Highly Reddened Quasars, ApJ, 607, 60
\bibitem[]{} Ivezic, Z.\ et al.\ (2002) Optical and Radio Properties of Extragalactic Sources Observed by the FIRST Survey and the Sloan Digital Sky Survey, AJ, 124, 2364 
\bibitem[]{} Kauffmann, G.\  \& Haehnelt, M.G.\ (2000) A unified model for the evolution of galaxies and quasars, MNRAS, 311, 576
\bibitem[]{} Kauffmann, G.\ et al.\ (2003) The host galaxies of active 
galactic nuclei, MNRAS, 346, 1055 
%\bibitem[]{} Kristian, J.\ (1973), Quasars as Events in the Nuclei of 
%Galaxies: the Evidence from Direct Photographs, ApJ, 179, L61
\bibitem[]{} Krolik, J.H., (2001) Systematic Errors in the Estimation of Black Hole Masses by Reverberation Mapping, ApJ, 551, 72
\bibitem[]{} Kukula, M.J., Dunlop, J.S., McLure, R.J., Miller, L., Percival, 
W.J., Baum, S.A.\ \& O'Dea, C.P.\ (2001) A NICMOS imaging study of high-z 
quasar host galaxies, MNRAS, 326, 1533
\bibitem[]{} Lacy, M., Laurent-Muehleisen, S.A., Ridgway, S.E., Becker, R.H.\ 
\& White, R.L.\ (2001) The Radio Luminosity-Black Hole Mass Correlation 
for Quasars from the FIRST Bright Quasar Survey and a ``Unification 
Scheme'' for Radio-loud and Radio-quiet Quasars, ApJ, 
551, L17 
\bibitem[]{} Lacy, M.\ et al.\ (2002), The Reddest Quasars. II. A Gravitationally Lensed FeLoBAL Quasar, AJ, 123, 2925
\bibitem[]{} Lacy, M., Gates, E.L., Ridgway, S.E., de Vries, 
W., Canalizo, G., Lloyd, J.P.\ \& Graham, J.R.\ (2002) Observations of 
Quasar Hosts with Adaptive Optics at Lick Observatory, AJ, 124, 3023 
\bibitem[]{} Lacy, M.\ et al.\ (2004), Obscured and Unobscured Active Galactic Nuclei in the Spitzer Space Telescope First Look Survey, ApJS, 154, 166
\bibitem[]{} Lacy, M.\ et al.\ (2005), Mid-infrared selection of Quasar-2s in Spitzer's First Look Survey, MmSAI, 76, 154
\bibitem[]{} Laor, A.\ (1998) On Quasar Masses and Quasar Host Galaxies, ApJ, 505, L83 
\bibitem[]{} Laor, A.\ (2000) On Black Hole Masses and Radio Loudness in Active Galactic Nuclei, ApJ, 543, L111
\bibitem[]{} Leipski, C.\ et al.\ (2005), The ISO-2MASS AGN survey: On the type-1 sources, A\&A, 440, L5
\bibitem[]{} Marble, A.R.\ et al.\ (2003), A Hubble Space Telescope WFPC2 Snapshot Survey of 2MASS-selected Red QSOs, ApJ, 590, 707
\bibitem[]{} Martinez-Sansigre, A.\ et al.\ (2005), Most supermassive black hole growth is obscured by dust, Nature, in press
\bibitem[]{} McLure, R.J.\ Kukula, M.J., Dunlop, J.S., Baum, S.A., O'Dea, C.P.\ \& Hughes, D.H.\ (1999) A comparative HST imaging study of the host galaxies of radio-quiet quasars, radio-loud quasars and radio galaxies - I, MNRAS, 
308, 377
\bibitem[]{} Mesinger, A.\ Haiman, Z.\ \& Cen, R.\ (2004) Probing the reionization history using the spectra of high-redshift sources, ApJ, 613, 23
\bibitem[]{} Metcalf, R.B.\ \& Magliocchetti, M.\ (2005) The role of black hole mass in quasar radio activity, MNRAS, submitted (astro-ph/0505194)
\bibitem[]{} Nelson, C.H., et al.\ (2004) The Relationship Between Black Hole Mass and Velocity Dispersion in Seyfert 1 Galaxies, ApJ, 615, 652
\bibitem[]{} Nelson, C.H.\ et al.\ (2000) ApJ, 544, L91
\bibitem[]{} Nolan, L.A.\ et al.\ (2001), The ages of quasar host galaxies, MNRAS, 323, 308
\bibitem[]{} Norman, C.\ et al.\ (2002) A Classic Type 2 QSO, ApJ, 571, 218
\bibitem[]{} Onken, C.A.\ et al.\ (2004), Supermassive Black Holes in Active Galactic Nuclei. II. Calibration of the Black Hole Mass-Velocity Dispersion Relationship for Active Galactic Nuclei, ApJ, 615, 645
\bibitem[]{} Padovani, P.\ et al.\ (2004), Discovery of optically faint obscured quasars with Virtual Observatory tools, A\&A, 424, 545
\bibitem[]{} Peterson, B.M.\ et al.\ (2004), Central Masses and Broad-Line Region Sizes of Active Galactic Nuclei. II. A Homogeneous Analysis of a Large Reverberation-Mapping Database, ApJ, 613, 682
\bibitem[]{} Richards, G.T.\ et al.\ (2001) Photometric Redshifts of 
Quasars, AJ, 121, 2308 
\bibitem[]{} Richards, G.T.\ et al.\ (2005) The 2dF-SDSS LRG and QSO (2SLAQ) Survey: the $z < 2.1$ quasar luminosity function from 5645 quasars to $g= 21.85$, 
MNRAS, 360, 839
\bibitem[]{} Ridgway, S.E., Heckman, T.M., Calzetti, D.\ \& Lehnert, M.\ 
(2001) NICMOS Imaging of the Host Galaxies of $z\sim 2-3$ Radio-quiet Quasars, 
ApJ, 550, 122
\bibitem[]{} Schneider, D.P.\ et al.\ (2003) The Sloan Digital Sky Survey Quasar Catalog. II. First Data Release, AJ, 126, 2579
\bibitem[]{} Shields, G.A.\ et al.\ (2003) The Black HoleBulge Relationship in Quasars, ApJ, 583, 124
\bibitem[]{} Silk, J.\ \& Rees, M.J.\ (1998) Quasars and galaxy formation, A\&A, 311, L1
%\bibitem[]{} Smith, E.P., Heckman, T.M., Bothun, G.D., Romanishin, W., 
%Balick, B.\ (1986) On the nature of QSO host galaxies, ApJ, 306, 64
\bibitem[]{} Soltan, A.\ (1982), Masses of quasars, MNRAS, 200, 115
\bibitem[]{} Stern, D.\ et al.\ (2002), Chandra Detection of a Type II Quasar at $z = 3.288$, ApJ, 568, 71
\bibitem[]{} Stockton, A.\ \& MacKenty, J.W.\ (1983) Extended emission-line regions around QSOs, Nat, 305, 678
\bibitem[]{} Stockton, A., Canalizo, G.\ \& Close, L.M.\ (1998) PG 1700+518 Revisited: Adaptive-Optics Imaging and a Revised Starburst Age for the Companion, ApJ, 500, L121
\bibitem[]{} Tadhunter, C.N.\ et al.\ 2005, Starbursts and the triggering of the activity in nearby powerful radio galaxies, MNRAS, 356, 480
\bibitem[]{} Vanden Berk, D.E.\ et al.\ (2001), Composite Quasar Spectra from the Sloan Digital Sky Survey, AJ, 122, 549
\bibitem[]{} Vanden Berk, D.E.\ et al.\ (2004),  The Ensemble Photometric Variability of $\sim $25,000 Quasars in the Sloan Digital Sky Survey, ApJ, 601, 692
\bibitem[]{} van der Marel, R.P. (1999), The Black Hole Mass Distribution in Early-Type Galaxies: Cusps in Hubble Space Telescope Photometry Interpreted through Adiabatic Black Hole Growth, AJ, 117, 744 
\bibitem[]{} Vaughan, S.\ \& Fabian, A.C.\ (2004) A long, hard look at MCG-6-30-15 with XMM-Newton II: detailed EPIC analysis and modelling, MNRAS, 348, 1415
\bibitem[] Vestergaard, M.\ (2002) Determining Central Black Hole Masses in Distant Active Galaxies, ApJ, 571, 733
\bibitem[]{} Wandel, A.\ (2002) Black Holes of Active and Quiescent Galaxies. I. The Black Hole-Bulge Relation Revisited, ApJ, 565, 762
\bibitem[]{} White, R.L.\ et al.\ (1999) The FIRST Bright Quasar Survey. II. 60 Nights and 1200 Spectra Later, 
ApJS, 126, 133
\bibitem[]{} White, R.L., Becker R.H., Fan, X.\ \& Strauss M.A. (2003)  Probing the Ionization State of the Universe at $z>6$, AJ, 126, 1
\bibitem[]{} Woo, J.-H.\ \& Urry, C.M.\ (2002) The Independence of Active Galactic Nucleus Black Hole Mass and Radio Loudness, ApJ, 581, L5
%\bibitem[]{} Wyckoff, S., Wehinger, P.A.\ \& Gehren, T.\ (1981)
%Resolution of quasar images, ApJ, 247, 750
\bibitem[]{} Yu, Q.\ \& Tremaine, S.\ (2002), Observational constraints on growth of massive black holes, 
MNRAS, 335, 965
\bibitem[]{} Zakamska, N., (2003) Candidate Type II Quasars from the Sloan Digital Sky Survey. I. Selection and Optical Properties of a Sample at $0.3<z<0.83$, AJ, 126, 2125

\end{thebibliography}

\printindex
\end{document}